\documentclass[aps,pra,letterpaper,twocolumn,showpacs,amsmath,amssymb,superscriptaddress]{revtex4-1}
\usepackage{graphicx}
\usepackage{setspace}
\usepackage{dcolumn}
\usepackage{mathrsfs}
\usepackage{bm}
\usepackage{longtable}

\usepackage{wrapfig,booktabs,lipsum}
\usepackage[title]{appendix}

\usepackage[colorinlistoftodos]{todonotes}
\definecolor{LightBlue}{rgb}{0.63, 0.79, 0.95}
\definecolor{LightGreen}{rgb}{0.5, 1.0, 0.0}
\definecolor{LightRose}{rgb}{0.98, 0.8, 0.91}
\definecolor{LightGray}{rgb}{0.75, 0.75, 0.75}
\definecolor{DeepPink}{rgb}{1.0, 0.08, 0.58}
\definecolor{Gold}{rgb}{1.0, 0.87, 0.0}
\definecolor{Beige}{rgb}{0.97, 0.91, 0.56}

\begin{document}

\preprint{}

\title{Measurement of the strontium triplet Rydberg series by depletion spectroscopy of ultracold atoms}

\author{Luc Couturier}

\affiliation{Shanghai Branch, National Laboratory for Physical Sciences at Microscale and Department of Modern Physics, University of Science and Technology of China, Hefei, Anhui 230026, China}

\affiliation{CAS Center for Excellence and Synergetic Innovation Center in Quantum Information and Quantum Physics, University of Science and Technology of China, Shanghai 201315, China}

\author{Ingo Nosske}

\affiliation{Shanghai Branch, National Laboratory for Physical Sciences at Microscale and Department of Modern Physics, University of Science and Technology of China, Hefei, Anhui 230026, China}

\affiliation{CAS Center for Excellence and Synergetic Innovation Center in Quantum Information and Quantum Physics, University of Science and Technology of China, Shanghai 201315, China}

\author{Fachao Hu}

\affiliation{Shanghai Branch, National Laboratory for Physical Sciences at Microscale and Department of Modern Physics, University of Science and Technology of China, Hefei, Anhui 230026, China}

\affiliation{CAS Center for Excellence and Synergetic Innovation Center in Quantum Information and Quantum Physics, University of Science and Technology of China, Shanghai 201315, China}

\author{Canzhu Tan}

\affiliation{Shanghai Branch, National Laboratory for Physical Sciences at Microscale and Department of Modern Physics, University of Science and Technology of China, Hefei, Anhui 230026, China}

\affiliation{CAS Center for Excellence and Synergetic Innovation Center in Quantum Information and Quantum Physics, University of Science and Technology of China, Shanghai 201315, China}

\author{Chang Qiao}

\affiliation{Shanghai Branch, National Laboratory for Physical Sciences at Microscale and Department of Modern Physics, University of Science and Technology of China, Hefei, Anhui 230026, China}

\affiliation{CAS Center for Excellence and Synergetic Innovation Center in Quantum Information and Quantum Physics, University of Science and Technology of China, Shanghai 201315, China}

\author{Y. H. Jiang}

\email{jiangyh@sari.ac.cn}

\affiliation{CAS Center for Excellence and Synergetic Innovation Center in Quantum Information and Quantum Physics, University of Science and Technology of China, Shanghai 201315, China}

\affiliation{Shanghai Advanced Research Institute, Chinese Academy of Sciences, Shanghai 201210, China}

\author{Peng Chen}

\email{peng07@ustc.edu.cn}

\affiliation{Shanghai Branch, National Laboratory for Physical Sciences at Microscale and Department of Modern Physics, University of Science and Technology of China, Hefei, Anhui 230026, China}

\affiliation{CAS Center for Excellence and Synergetic Innovation Center in Quantum Information and Quantum Physics, University of Science and Technology of China, Shanghai 201315, China}

\author{Matthias Weidem\"uller}

\email{weidemueller@uni-heidelberg.de}

\affiliation{Shanghai Branch, National Laboratory for Physical Sciences at Microscale and Department of Modern Physics, University of Science and Technology of China, Hefei, Anhui 230026, China}

\affiliation{CAS Center for Excellence and Synergetic Innovation Center in Quantum Information and Quantum Physics, University of Science and Technology of China, Shanghai 201315, China}

\affiliation{Physikalisches Institut, Universit\"at Heidelberg, Im Neuenheimer Feld 226, 69120 Heidelberg, Germany}

 \date{\today}

\begin{abstract}
We report on the atom loss spectroscopy of strontium Rydberg atoms in a magneto-optical trap, using a two-photon excitation scheme through the intermediate state $5\mathrm{s}5\mathrm{p} \, ^3\mathrm{P}_1$. Energies of the $5\mathrm{s}n\mathrm{s} \, ^3\mathrm{S}_1$ and $5\mathrm{s}n\mathrm{d} \, ^3\mathrm{D}_{1,2}$ Rydberg series of $^{88}$Sr in the range $13 \leq n \leq 50$ are determined with an absolute accuracy of 10 MHz, including the perturbed region where the $5\mathrm{s}n\mathrm{d} \, ^3\mathrm{D}_{2}$ series couples to the $5\mathrm{s}n\mathrm{d} \, ^1\mathrm{D}_{2}$ series. This represents an improvement by more than two orders of magnitude compared to previously published data. The quantum defects for each series are determined using the extended Rydberg-Ritz formula in the range where there is no strong perturbation. A value of $1\,377\,012\,721(10) \,\mathrm{MHz}$ for the first ionization limit of $^{88}$Sr is extracted.

\end{abstract}


\maketitle


\section{Introduction}

Spectroscopy of the intricate electronic level structure of two-electron Rydberg atoms has stimulated the development of theoretical models for the description of correlated electrons, in particular multi-channel quantum defect theory (MQDT) \cite{Seaton1983,Aymar1996}. The interaction between the electrons leads to phenomena such as, e.g., autoionization of Rydberg states \cite{Cooke1978,Poirier1988}. With the advent of laser cooling techniques, new opportunities for applications of two-electron Rydberg atoms were identified \cite{Dunning2016}. The optical transition of the ion core of the Rydberg atoms allows for dipolar trapping of Rydberg atoms in lattices \cite{Mukherjee2011} but also new techniques such as spatially selective Rydberg atom detection through autoionization \cite{Lochead2013}. Alkaline-earth like atoms feature narrow intercombination lines, which, in combination with strong Rydberg-Rydberg interaction, make them a good candidate for quantum simulation via Rydberg dressing \cite{Gaul2016,bounds2018} and for the generation of spin-squeezed states with applications in metrology \cite{gil2014}.

In the case of strontium, the singlet Rydberg series are well known since the early days of laser spectroscopy \cite{Beigang1982a} and have been further studied over the past years \cite{mauger2007,millen2010}. Rydberg excitation of the triplet Rydberg states in an ultracold atomic gas has been realized only recently \cite{bridge2016, DeSalvo2016, camargo2018, bounds2018}. Surprisingly, available spectroscopic data about the triplet series date back to the late 70's \cite{Esherick1977, armstrong1979, Beigang1982}, where the measurements were performed with hot atoms at pressures on the order of $\sim10^{-2}\,\mathrm{mbar}$, leading to a significant line broadening. Accuracies of the absolute transition frequencies are on the order of few GHz. Such a large uncertainty has been the limiting factor in predicting properties of the strontium triplet series \cite{vaillant2012,vaillant2014}.

In this paper we present spectroscopic data with an improved accuracy of the triplet Rydberg series $5\mathrm{s}n\mathrm{s}\,^3\mathrm{S}_1$, $5\mathrm{s}n\mathrm{d}\,^3\mathrm{D}_1$ and $5\mathrm{s}n\mathrm{d}\,^3\mathrm{D}_2$ in the range $n=13...50$. The data include the strongly perturbed region of the $5\mathrm{s}n\mathrm{d}\,^3\mathrm{D}_2$ series which couples to the singlet series. The measurement of the triplet series is performed on an ultracold gas of $^{88}$Sr atoms using standard spectroscopic techniques \cite{Hostetter2015, Lehec2018}. The Rydberg transitions are detected through atom-loss spectroscopy in a magneto-optical trap operated on the Sr intercombination line. The spectroscopic data are fitted to the extended Rydberg-Ritz formula far away from perturbations to extract reliable quantum defects for energy level prediction. We also extract an updated value for the first ionization limit of $\mathrm{^{88}Sr}$ \cite{Beigang1982a}.


\section{Experimental methods}
\label{sec:expsetup}

\subsection{Rydberg excitation}

\begin{figure}
\includegraphics[angle=-90,width=\linewidth]{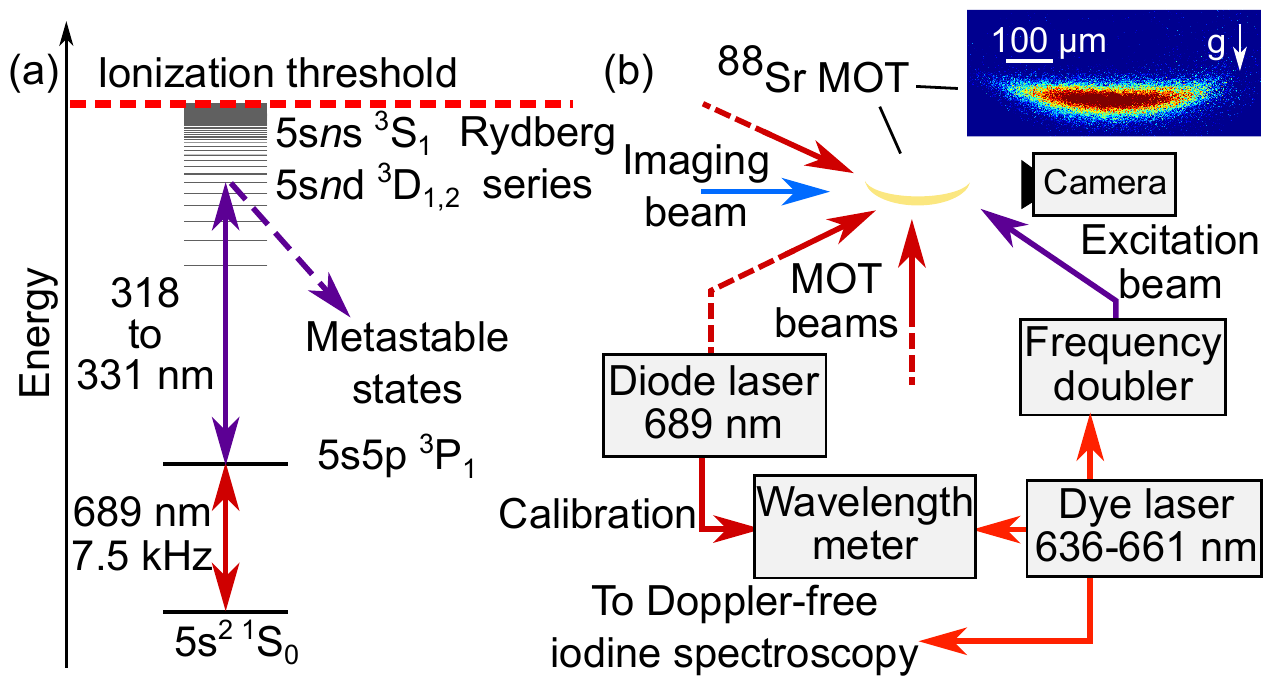}
\caption{Rydberg excitation scheme. (a) Energy level scheme for strontium Rydberg excitation. The atoms are excited to the $5\mathrm{s}n\mathrm{s}\,^3\mathrm{S}_1$, $5\mathrm{s}n\mathrm{d}\,^3\mathrm{D}_1$ and $5\mathrm{s}n\mathrm{d}\,^3\mathrm{D}_2$ by a two-photon excitation via the $5\mathrm{s}5\mathrm{p}\,^3\mathrm{P}_1$ state. The UV light is tunable from $n=13$ to the first ionization limit. The metastable states are the long-lived $5\mathrm{s}5\mathrm{p}\,^3\mathrm{P}_0$ and $5\mathrm{s}5\mathrm{p}\,^3\mathrm{P}_2$ to which Rydberg state decay through direct or cascade deexcitation. (b) Schematic of the experiment. A Magneto-optical trap (MOT) is operated on the $5\mathrm{s}^2\,^1\mathrm{S}_0\rightarrow 5\mathrm{s}5\mathrm{p}\,^3\mathrm{P}_1$ transition at 689 nm from which atoms are excited by a frequency doubled dye laser. The MOT atom number is monitored by absorption imaging and a typical optical density map is shown in the upper right corner. An iodine saturated absorption spectroscopy is used to determine the accuracy of the wavelength meter.
}
\label{fig:ExpSetup}
\end{figure}

A strontium magneto-optical trap (MOT), operated on the $5\mathrm{s}^2\,^1\mathrm{S}_0\rightarrow \mathrm{5s5p}\,^1\mathrm{P}_1$ transition, is loaded from a strontium two-dimensional MOT as described in Ref.~\cite{nosske2017}. Atoms are then transferred to a MOT operated on the narrow transition $\mathrm{5s^2}\,^1\mathrm{S}_0\rightarrow \mathrm{5s5p}\,^3\mathrm{P}_1$ transition (see, e.g., \cite{Stellmer2014}), at a magnetic field gradient of $5\, \mathrm{G}/\mathrm{cm}$, as shown schematically in Fig. \ref{fig:ExpSetup} (b). We obtain about $10^6$ atoms at a temperature of $1\,\mathrm{\mu K}$. Due to the narrow linewidth of the transition, the atoms accumulate in the lower shell of an ellipsoid, as shown in the inset of Fig. \ref{fig:ExpSetup} (b), with a peak atomic density about $2\times 10^{10}\,\mathrm{atoms/cm^3}$ \cite{Hanley2017}. The shape of the atomic cloud reflects the fact that the Zeeman shift compensates the MOT beam detuning, as described in Refs. \cite{Hanley2017,Loftus2004}, at a finite magnetic field offset, essentially pointing along the vertical axis.

We excite the atoms to Rydberg states with two photons using the transitions $5\mathrm{s}^2\,^1\mathrm{S}_0\rightarrow 5\mathrm{s}5\mathrm{p}\,^3\mathrm{P}_1\rightarrow 5\mathrm{s}nl\,^3L_J$, as shown in Fig. \ref{fig:ExpSetup} (a). Starting from the $5\mathrm{s}5\mathrm{p}\,^3\mathrm{P}_1$ state, we can access the $5\mathrm{s}n\mathrm{s}\,^3\mathrm{S}_1$, $5\mathrm{s}n\mathrm{d}\,^3\mathrm{D}_1$ and $5\mathrm{s}n\mathrm{d}\,^3\mathrm{D}_2$ Rydberg series (abbreviated $^3\mathrm{S}_1$, $^3\mathrm{D}_{1,2}$ in the rest of the paper). These states decay back to $\mathrm{5s5p}\,^3\mathrm{P}_1$ but also $\mathrm{5s5p}\,^3\mathrm{P}_2$ and $\mathrm{5s5p}\,^3\mathrm{P}_0$ which are long-lived metastable states. The first photon is provided by the MOT laser field (see Fig \ref{fig:ExpSetup} (a)), which is generated by a 689 nm diode laser. The MOT is operated at a saturation parameter $s\approx 20$. The laser is stabilized to an ultrastable cavity reducing its linewidth to less than 10 kHz. The cavity drift amounts to a $8\,\mathrm{kHz}/\mathrm{day}$ laser frequency deviation which is compensated by using saturated absorption spectroscopy in a strontium heat pipe as a reference.

We excite the atoms in the $\mathrm{5s5p}\,^3\mathrm{P}_1,m_J=+1$ state to a Rydberg state with a UV beam of $1.4\,\mathrm{mm}$ $1/e^2$ diameter, larger than the size of the MOT. The UV beam is linearly polarized along the vertical direction, which drives $\mathrm{\pi}$ transitions due to the magnetic alignment of the atoms in the narrow-band MOT \cite{Hanley2017,Loftus2004}. We use a frequency doubled dye laser which can be tuned from $\lambda=318\,\mathrm{nm}$ to $\lambda=331\,\mathrm{nm}$, in a setup similar to the one described in Ref. \cite{arias2017}. The UV laser has a linewidth below $200\,\mathrm{kHz}$ over $100\,\mathrm{ms}$. We use a UV pulse of one to few ms, with a power from few tens of $\mathrm{\mu W}$ to few $\mathrm{mW}$, adjusted to keep a reasonable contrast as the loss is observed to increase at lower $n$.

The UV laser frequency is scanned over the transition twice in each direction at a scan speed of $\sim200\,\mathrm{kHz/s}$. Atoms decaying to the metastable states, through direct or cascade decay induced by blackbody radiation, do not participate to the cooling cycle any longer and result in atom loss when performing absorption imaging at $461\,$nm on the $\mathrm{5s}^2\,^1\mathrm{S}_0\rightarrow \mathrm{5s5p}\,^1\mathrm{P}_1$ transition. We determine the number of remaining atoms before and after the Rydberg excitation. The overall repetition rate for the detection of Rydberg atoms is $0.5\,\mathrm{Hz}$. There is a finite loss of $\sim30\,\%$ occurring without Rydberg excitation due to the finite MOT storage time. We plot the rescaled atom number versus the total energy, as shown in Fig. \ref{fig:single_RyLine}, for each Rydberg line.

\subsection{Determination of the energy levels}

The total energy is deduced from the sum of the two photon energies, at $689\,\mathrm{nm}$ and $318...331\,\mathrm{nm}$. The energy of the first photon corresponds to the literature value of the transition for the $\mathrm{5s5p\,^3P_1}$ state \cite{Ferrari2003} plus a finite detuning of $\Delta f_{\mathrm{MOT}}=-600\,\mathrm{kHz}$ corresponding to the MOT laser detuning. This detuning is known on a $10\,\mathrm{kHz}$ level through absorption spectroscopy in a heat pipe.

As shown in Fig. \ref{fig:ExpSetup} (b), we determine the frequency of the Rydberg excitation beam by measuring the frequency $ f_{\mathrm{dye}}^{\mathrm{WLM}}$ of the Rydberg excitation laser with a commercial wavelength meter (HighFinesse WSU-10). The wavelength meter has a specified accuracy of $10\,\mathrm{MHz}$ at three standard deviation for a range of $\pm200\,\mathrm{nm}$ around the calibration wavelength. We calibrate the wavelength meter with the $689\,\mathrm{nm}$ laser, for which the corresponding strontium resonance frequency $f_{689}^{\mathrm{lit}}$ is known to an accuracy of $10\,\mathrm{kHz}$.

As an additional frequency calibration close to the respective Rydberg lines, part of the light of the dye laser is sent to a saturated absorption spectroscopy of iodine. As described in Appendix \ref{App:AbsFreq} we find a systematic frequency shift of $\delta f_{\mathrm{sys}}^{\mathrm{WLM}}=16.8\,\textrm{MHz}$ with a statistical error of $\pm9.4\,\textrm{MHz}$ (at $1\,\sigma$) on the frequency reading. To determine the Rydberg state energy,  we also include the Zeeman shift $\delta f_{\mathrm{ZS}}^{\mathrm{Ry}}$ of the Rydberg states, which is series dependent and typically $|\delta f_{\mathrm{ZS}}^{\mathrm{Ry}}|<500\,\textrm{kHz}$. 

\begin{figure}
\includegraphics[width=\linewidth]{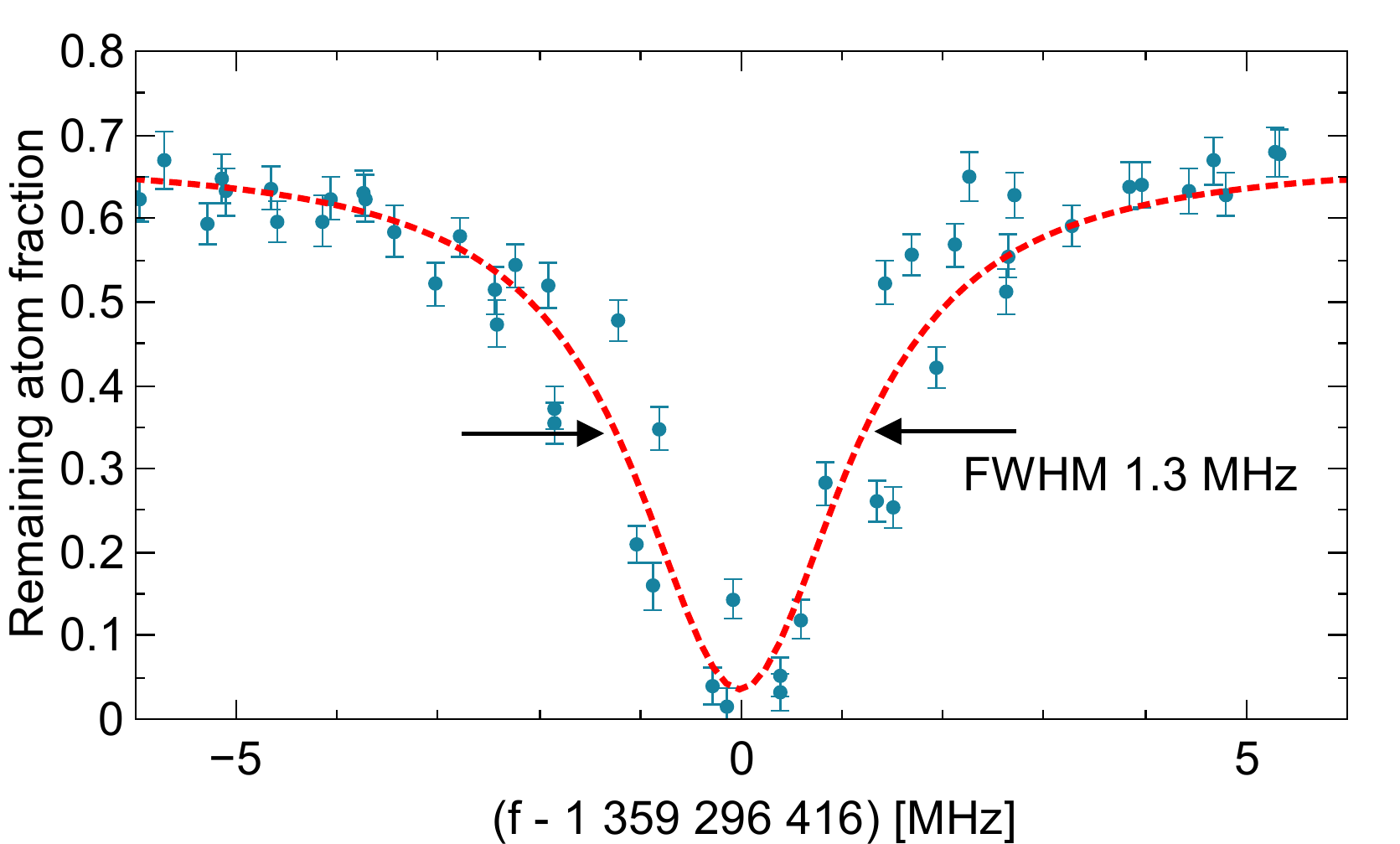}
\caption{Atom loss spectroscopy of the Rydberg line $5\mathrm{s}17\mathrm{s}\,^3\mathrm{S}_1$. The blue circles is the fraction of atom number measured by absorption imaging remaining after Rydberg excitation in the MOT. The red dashed line is a Lorentzian fit. The frequency axis is the relative UV laser frequency with the origin set at the fitted center frequency. The error bars are estimated by analyzing the noise on the area of the imaging pictures where no atom is present.}
\label{fig:single_RyLine}
\end{figure}

To determine the center of the Rydberg lines, we fit the spectrum obtained by scanning the UV light frequency with a Lorentzian function, as exemplarily shown in Fig. \ref{fig:single_RyLine}.
The full-width half-maximum is typically $1\,\textrm{MHz}$ and the fit error on the center determination from the fit is typically $10\,\textrm{kHz}$.

The uncertainty  from the wavelength meter reading as described above is by far the major contribution to the uncertainty. Due to the small Rabi frequencies (below 500 kHz), AC Stark shifts are negligible on this scale. The frequency shifts due to interactions are limited by the excitation linewidth, i.e. the laser linewidth of $200\,\textrm{kHz}$ or the Rabi frequency, which is much smaller than the wavelength meter uncertainty. To our knowledge, the only published measurements of the DC polarizabilities for the triplet Rydberg states are found in Refs. \cite{camargo2017,bounds2018t}. Using the rescaled polarizability, a DC Stark shift of $100\,\mathrm{kHz}$ would correspond to a stray field of $30\,\mathrm{mV/cm}$ for the $5\mathrm{s}50\mathrm{d}\,^3\mathrm{D}_1$ state, which is the measured state with one of the highest polarizabilities. Considering that there is no electrode inside or outside the steel vacuum chamber, these residual electric field values are assumed to be small and we therefore neglect a residual DC Stark shift. This is confirmed by the fact that  there is no observable contribution in the Rydberg state energies which scales as $\propto n^{*7}$ like the polarizability. When added in quadrature, we obtain a total uncertainty of $10\,\mathrm{MHz}$ (rounded to a $1\,\mathrm{MHz}$ precision).


\section{Results and discussion}
\label{sec:results}

\begin{figure}
\includegraphics[width=\linewidth]{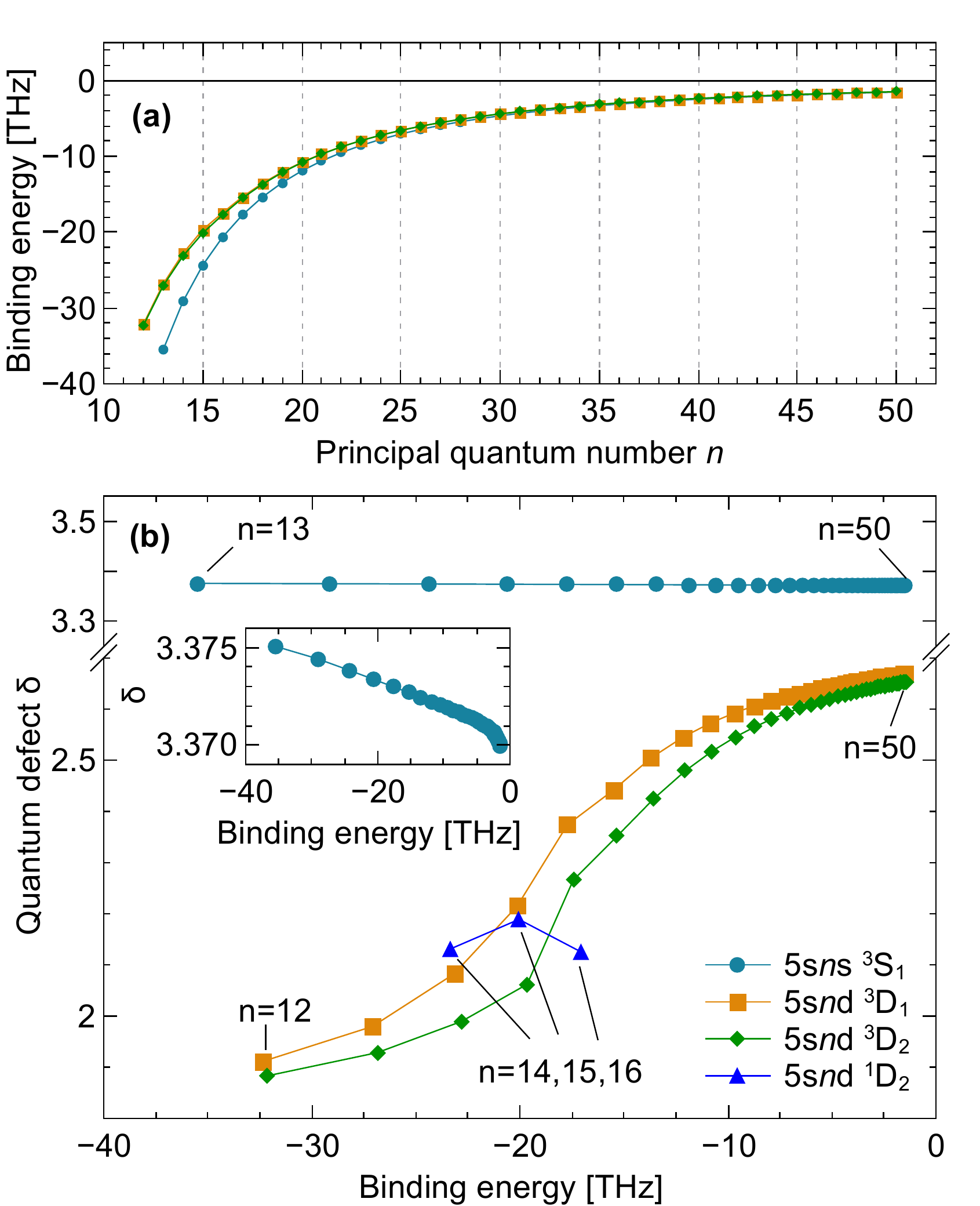}
\caption{Experimental states energies and quantum defects of the $5\mathrm{s}n\mathrm{s}\,^3\mathrm{S}_1$and  $5\mathrm{s}n\mathrm{d}\,^3\mathrm{D}_{1,2}$ Rydberg series. (a) Measured binding energies of the triplet Rydberg series. The exact values can be found in the tables \ref{tab:SseriesEnergyTable} and \ref{tab:DseriesEnergyTable} in the Appendix \ref{app:RyEnergies}. (b) Quantum defects deduced from the Rydberg state energies by the Rydberg-Ritz formula versus its binding energy. The perturbation around $n=15$ creates a coupling between $5\mathrm{s}n\mathrm{d}\,^3\mathrm{D}_2$ series with the singlet series $5\mathrm{s}n\mathrm{d}\,^1\mathrm{D}_2$. The solid lines are guides to the eye. The inset is a zoom on the $5\mathrm{s}n\mathrm{s}\,^3\mathrm{S}_1$ Rydberg series which exhibits a slight energy dependence. The error bars are much smaller than the symbols.}
\label{fig:lufanoplot}
\end{figure}

Unlike the singlet Rydberg series which have been determined with a an accuracy of $30\,\mathrm{MHz}$ \cite{Beigang1982a,rubbmark1978}, the triplet series $^3\mathrm{S}_1$ and $^3\mathrm{D}_{1,2,3}$ have only been measured previously with an accuracy on the order of few gigahertz \cite{Beigang1982,Esherick1977}. 
With our setup we have improved the accuracy on the transition frequencies to $10\,\mathrm{MHz}$, i.e. two orders of magnitude, for the $^3\mathrm{S}_1$ and $^3\mathrm{D}_{1,2}$ triplet Rydberg series which are accessible through dipole transitions. We have mapped out all transition energies from $n = 13$ to $n=50$ for these series; they are plotted in Fig. \ref{fig:lufanoplot}(a). The measured values are given by the Tables \ref{tab:SseriesEnergyTable} and \ref{tab:DseriesEnergyTable} in Appendix \ref{app:RyEnergies}. 

The energies can be described by the Rydberg-Ritz formula:
\begin{equation}
 E_n= I_s-\frac{\tilde{R}}{(n-\delta(n))^2},
 \label{eqn:RyRitz}
\end{equation}
where $I_s$ is the first ionization threshold, $\tilde{R}$ is the mass-corrected Rydberg constant for $\mathrm{^{88}Sr}$, $n$ is the principal quantum number and $\delta(n)$ is the quantum defect, which is specific to each Rydberg series. $\tilde{R}$ is taken as $109\,736.631\,\mathrm{cm^{-1}}$ using the latest values of the fundamental constants \footnote{In all the previous literature, the value of the Rydberg constant was taken as  $109\,736.627\,\mathrm{cm^{-1}}$, calculated in \cite{rubbmark1978}, which uses older values of the fundamental constants} and of the strontium mass \cite{DeLaeter2003,mohr2016}. An accurate prediction of $\delta(n)$ using a model allows to reproduce and predict the Rydberg energies.

In Fig. \ref{fig:lufanoplot}(b) we show a plot of the experimental quantum defects versus the binding energy of the Rydberg states of the different series. It shows the energy dependence of the quantum defect and perturbations of the series, as described in detail in previous works \cite{vaillant2014,Beigang1983}. Proper description of these energies would require MQDT\cite{Seaton1983,Aymar1996}, which is beyond the scope of this paper. We can however extract some qualitative features. For small binding energies, the energy levels converge to the ionization energy, as expected from Eq. \ref{eqn:RyRitz}.

In the case of the $^3\mathrm{S}_1$ series, the quantum defect is nearly independent of the binding energy, indicating a small influence of the ionic core polarizability. The small residual energy dependence is depicted by the inset in Fig. \ref{fig:lufanoplot}(b), which was not resolved in previous work \cite{Beigang1983}. The $^3\mathrm{D}_1$ series is strongly perturbed near $n=15$, that is attributed to a coupling to the $^3\mathrm{D}_3$ series \cite{vaillant2014}. These perturbations for two-electron Rydberg atoms are essentially due to admixtures of doubly-excited states which shift the position of the Rydberg level. The $^3\mathrm{D}_2$ series also exhibits a similar perturbation around $n=15$ . The behavior was explained by a six-channel MQDT \cite{vaillant2014}, which includes a coupling to the $^1\mathrm{D}_2$ series through a doubly excited state. Through the admixture, the transition from the $\mathrm{5s5p\,^3P_1}$ to the $^1\mathrm{D}_2$ series becomes dipole allowed. As a consequence, we can observe three states of this series for $n=14,15,16$. We assign the lines according to Refs. \cite{Beigang1982a,Beigang1982}, even though they are not pure states due to the strong mixing described in Ref. \cite{vaillant2014} . 

\begin{figure}
\includegraphics[width=\linewidth]{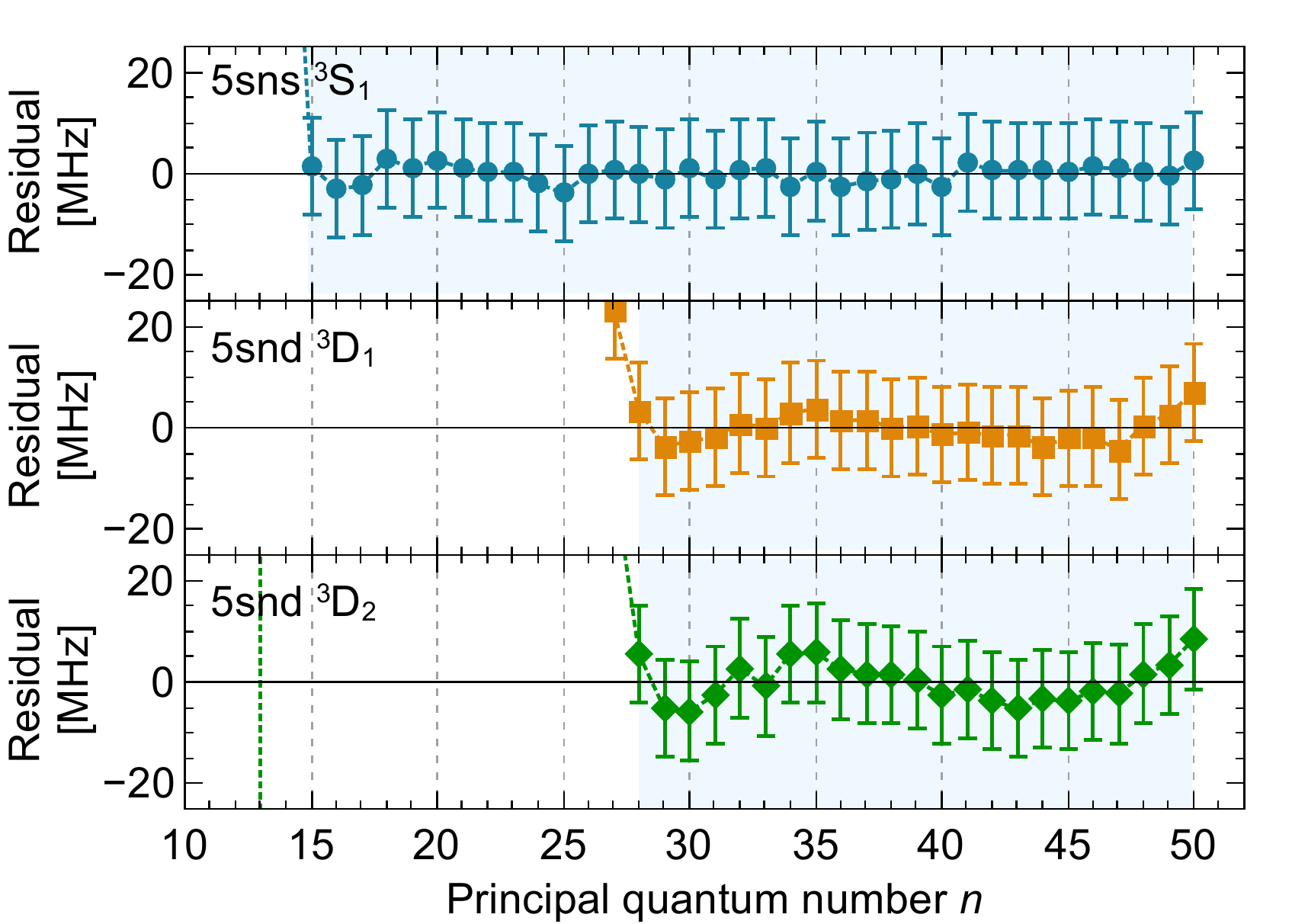}
\caption{Residuals of a fit of the experimental data for the $5\mathrm{s}n\mathrm{s}\,^3\mathrm{S}_1$ and $5\mathrm{s}n\mathrm{d}\,^3\mathrm{D}_{1,2}$ Rydberg series, respectively, with the extended Rydberg-Ritz formula. The results of the fits are given in Table \ref{tab:QuantumDefects}. The shaded background is the fitting range which has been optimized to minimize the error on the fitting parameters.}
\label{fig:residue}
\end{figure}

For practical purposes, we  perform an analysis far away from the perturbation of the Rydberg series and describe the quantum defect using the extended Rydberg-Ritz formula:
\begin{equation}
 \delta(n)=\delta_0 +\frac{\delta_2}{(n-\delta_0)^2}+\frac{\delta_4}{(n-\delta_0)^4}+...
 \label{eqn:QD}
\end{equation}
with $\delta_i$ ($i=0,2,4...$) parameters that have to be extracted from a fit to the experimental data. We perform such a fit of Eq. \ref{eqn:RyRitz} combined with Eq. \ref{eqn:QD}, with $\delta_i$ and the ionization limit $I_s$ as free parameters. We choose the fitting range such that the standard error on the fitted parameters is minimized. The fit results are shown in Table \ref{tab:QuantumDefects}, and the residuals of the fits are depicted in Fig. \ref{fig:residue} for the series $^3\mathrm{S}_1$, $^3\mathrm{D}_1$ and $^3\mathrm{D}_2$. All three series can be well reproduced within the selected fitting range by including up to $\delta_4$. Higher orders do not improve the quality of the fit. Our findings represent a one to three order of magnitude improvement on the evaluation of the first term of the quantum defect $\delta_0$, as compared to Ref. \cite{vaillant2012}. The improved coefficients can be used to extrapolate the Rydberg state energies at higher $n$. However at lower principal quantum number, there are significant deviations due to Rydberg series perturbations and would require a MQDT model for an accurate description.

\begin{table*}[t]
  \centering\small
  \begin{tabular}{|l l r  l l l|}
  \hline Series &\hspace{15pt} $\delta_0$\hspace{20pt} & $\delta_2\hspace{10pt}$\hspace{5pt}  & \hspace{10pt}$\delta_4$ & $I_S$ (MHz) &  Fitted range \\ \hline
  $5\mathrm{s}n\mathrm{s} \, ^3\mathrm{S}_1$ & 3.370\,778(4) & 0.418(1) & $-0.3(1)$ & 1\,377\,012\,720.6(7)& $15 \leq n \leq 50$\\
  $5\mathrm{s}n\mathrm{d} \, ^3\mathrm{D}_1$ & 2.675\,17(20) & -13.15(26) &$-4.444(91)\times 10^3$ & 1\,377\,012\,718(8)&$28 \leq n \leq 50$\\
  $5\mathrm{s}n\mathrm{d} \, ^3\mathrm{D}_2$ & 2.661\,42(30) &$-16.77(38)$ &$-6.656(134)\times 10^3$ &1\,377\,012\,718(12)& $28 \leq n \leq 50$\\
  \hline
  \end{tabular}\\
  \caption{Fitted quantum defects parameters  $\delta_k$ ($k=0,2,4$) and the ionization limit $I_S$ according to Eq. \ref{eqn:RyRitz} and \ref{eqn:QD}. The fitted range has been optimized to minimize the residual at high n, even though the series cannot be described fully by the Rydberg-Ritz formula. The uncertainties are obtained from the fit and are larger than the precision needed to reproduce the experimental data on a MHz level. }
  \label{tab:QuantumDefects}
\end{table*}

The ionization limit is determined from independent fits of the three Rydberg series (see Table \ref{tab:QuantumDefects}). All three values agree with each other within the error bar. We calculate a mean value weighted by the inverse of the square of the errors. The error on the ionization limit is taken as the uncertainty on the experimental data. The ionization limit for strontium $\mathrm{^{88}Sr}$ is thus  $1\,377\,012\,721(10) \,\mathrm{MHz}$. This value is $62\,\mathrm{MHz}$ higher than the value from Ref. \cite{Beigang1982a}. The discrepancy can be explained by the lower $n$ range used in the original work to extract the ionization limit, which is subject to Rydberg series perturbation.


\section{Conclusion}

We have measured the total energy of the strontium Rydberg states for the $^3\mathrm{S}_1$, $^3\mathrm{D}_1$ and $^3\mathrm{D}_2$ Rydberg series over the range of $n = 13$ to $n=50$ by depletion spectroscopy in a magneto-optical trap operated on the narrow intercombination line, yielding spectral lines with a linewidth around $1\,\mathrm{MHz}$. Using the precision of a high precision wavelength meter combined with the absolute accuracy of an iodine absorption spectroscopy, we have achieved a $10\,\mathrm{MHz}$ accuracy on the determination of the Rydberg energy levels and of the ionization limit. The improvement of the accuracy on these energies, in particular in the strongly perturbed region of the spectra, will be useful for improved theoretical predictions of the energy level positions \cite{vaillant2014} and of the Rydberg-Rydberg interactions \cite{vaillant2012}, which can be in turn used to predict more accurately more complex effects such as Rydberg dressing \cite{Gaul2016}. The existence of a considerable mixing between the $^3\mathrm{D}_2$ and $^1\mathrm{D}_2$ Rydberg series around $n=15$  has been confirmed through the direct observation of singlet states, which might offer interesting perspectives for optical multiwave mixing via Rydberg singlet-triplet coupling. 

To further improve on the energy determination, one would need major improvements on the experimental setup. So far, the determination of the energy levels is mainly limited by the absolute accuracy of the standard wavelength meter combined with a simple iodine spectroscopy, but could be greatly reduce by the use of a frequency comb \cite{Kliese2016}. At this level, one would need a trap-free measurement with an accurate electric and magnetic field control by having electrodes and additional coils, that would result in even narrower Rydberg lines.

\paragraph*{Note Added}:
Recently, the group of F. B. Dunning and T.C. Killian has presented data and analysis on the spectroscopy of triplet Rydberg series of $\mathrm{^{87}Sr}$ at high principal quantum numbers \cite{Ding2018-arxiv}. They make use of previously published data for $\mathrm{^{88}Sr}$ to evaluate the hyperfine splitting of $\mathrm{^{87}Sr}$ Rydberg states and use $\mathrm{^{87}Sr}$ energy measurements to improve quantum defect predictions for $\mathrm{^{88}Sr}$. From an estimation of the first ionization limit, they conclude that the previous value of the first ionization published in \cite{Beigang1982a} has to be shifted to higher energy, in agreement with our findings.


\section*{Acknowledgements}

We would like to thank Eberhard Tiemann for providing the \textit{IodineSpec} software, which allowed us to extract absolute transitions frequencies from the iodine spectra. M.W.'s research activities in China are supported by the 1000-Talent-Program of the Chinese Academy of Science. The work was supported by the National Natural Science Foundation of China (Grant Nos. 11574290 and
11604324) and Shanghai Natural Science Foundation (Grant No. 18ZR1443800). Y.H.J. also acknowledges support under Grant Nos. 11420101003 and 91636105.
P.C. acknowledges support of Youth Innovation Promotion Association, CAS.


\appendix

\section{Wavelength meter calibration by iodine spectroscopy}
\label{App:AbsFreq}

\begin{figure}
\includegraphics[width=\linewidth]{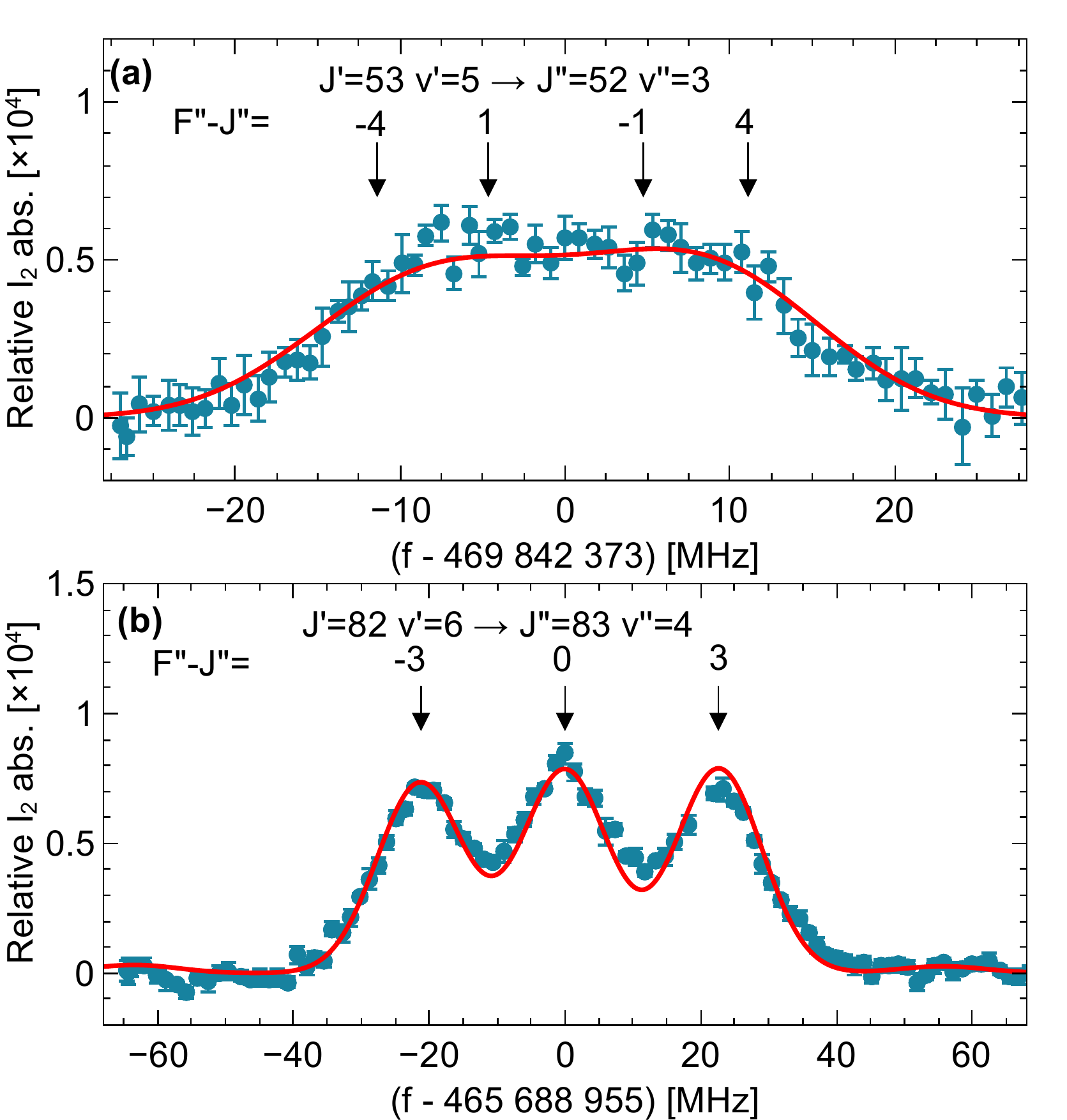}
\caption{Iodine Doppler-free spectrum. (a) and (b) Two examples of Doppler-free saturation spectroscopy of iodine lines near $638\,\mathrm{nm}$ and $644\,\mathrm{nm}$. The $y$ axis is the relative absorption through the iodine cell.  The blue circles are the experimental data. The red solid line is spectrum generated by computer software \cite{Bodermann2002}, rescaled and adjusted to the data by a fit, see text for details. The black arrows are the positions of the hyperfine lines as predicted by the theory. $J'$ and $J''$  are the lower and upper rotational quantum number. $\nu'$ and $\nu''$  are the lower and upper vibrational quantum number and $F$ is the quantum number of the total momentum of the upper level.}
\label{fig:I2}
\end{figure}

To have a reliable estimate of the uncertainty of the frequency measurement by the wavelength meter, we employ Doppler-free spectroscopy of iodine, which has a well-known spectrum \cite{Lukashov2018}. The Doppler-free hyperfine lines have a finite width and partially overlap with each other, as shown by two typical spectra depicted in Fig. \ref{fig:I2}. For a more accurate frequency determination, we chose groups of hyperfine lines with a comparatively small width. 

To determine the frequency, we simulate the spectrum $S_{I_2,sim}(f)$ with the software \textit{IodineSpec} \cite{I2Spec,Bodermann2002} which provides an absolute accuracy about $\pm 1.5\,\mathrm{MHz}$ (at $1\,\sigma$). We adjust the entire spectrum by fitting the parameters $a$ and $\delta_f$ such that the experimental data overlap with $S_{I_2,adj}(f)=a\times S_{I_2,sim}(f+\delta_f)$, as plotted in \ref{fig:I2} (a) and (b) with a red solid line.

\begin{figure}
\includegraphics[width=\linewidth]{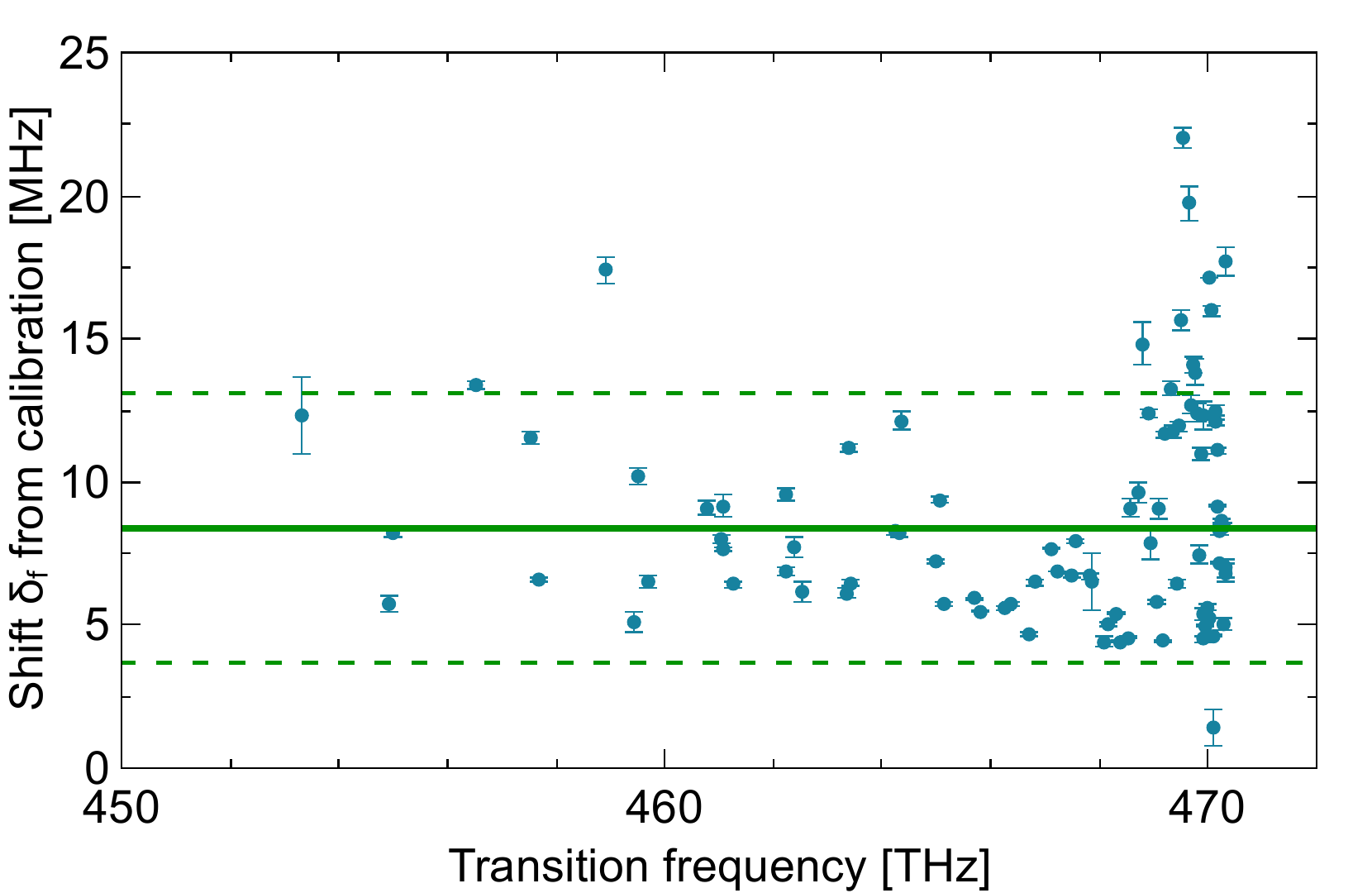}
\caption{Frequency shift $\mathrm{\delta_f}$ of the iodine spectra from the value computed by the \textit{IodineSpec} software \cite{I2Spec} for different $I_2$ lines at different frequencies. Each shift $\delta_f$ is extracted from a fit with the experimental data, as described in the text. The green solid line corresponds to the weighted mean shift. The dashed green lines corresponds to $\pm1$ standard deviation from the weighted mean.}
\label{fig:I2_stats}
\end{figure}

We choose iodine lines close to Rydberg resonances. We thus obtain 91 absolute frequencies whose position in frequency follows the Rydberg spectrum. Fig. \ref{fig:I2_stats} shows the deviation of the iodine line position of the spectra, for which the frequency is acquired by the wavelength meter, with the value from the \textit{IodineSpec} software. As there is no obvious trend in this shift as a function of the transition frequency, we estimate the shift by the statistical mean of all measurements weighted by their respective error bar. We find a mean value of $\delta_f$ of $+8.4\,\mathrm{MHz}$ with a standard deviation of $4.7\,\mathrm{MHz}$, which have to be multiplied by two for the UV frequency after frequency doubling. The standard deviation is used as a statistical error of the wavelength meter reading, even though the distribution is not Gaussian.  Following the statistical analysis, $85\,\%$ of the data points fall into one standard deviation, which indicates that the standard deviation might actually overestimate the real error.

The statistical uncertainty includes three sources of error: 1) The fitting error of the calculated spectrum to the data, 2) the statistical uncertainties of the spectrum predicted by the calculation, 3) the statistical error of wavelength measurement itself, which includes a possible long-term drift of the wavelength meter as the lines have been measured over a ten day period. The first source of error is the largest as our experimental data for the iodine spectrum have large error bars due to the electronic noise and distortion of the absorption signals which affects the fitting procedure. The second source of uncertainty is expected to be around $1.5\,\mathrm{MHz}$, that is, the statistical error of the original data used by the software. As for the last source of error, from a previous work in Ref. \cite{Couturier2018}, we have shown that this wavelength meter has a relative reading stability of $1.4\,\mathrm{MHz}$ at a $1\,\sigma$ level over 10 hours (taking the frequency doubling into account). All three sources of error contribute to the statistical error that we provide as the error bar of the wavelength meter.

\section{Experimental energies of the Rydberg states}
\label{app:RyEnergies}

The experimental energies of the Rydberg states for the $5\mathrm{s}n\mathrm{s} \, ^3\mathrm{S}_1$ and $5\mathrm{s}n\mathrm{d} \, ^3\mathrm{D}_{1,2}$ Rydberg series are presented in Tables \ref{tab:SseriesEnergyTable} and \ref{tab:DseriesEnergyTable}. The total energies are referenced to the ground state $\mathrm{5s}^2\,\mathrm{^1S_0}$ and corrected for the systematic frequency shifts as presented in the main text.

\begin{table*}[!h]
 \small
\centering
  \begin{tabular}{|c c c  || c c c |}
  \hline 
  $n$ & Series & $E_{exp}$ (MHz) & $n$ & Series & $E_{exp}$ (MHz) \\ 
  \hline
13 & $^3S_1$ & 1\,341\,500\,517 & 32 & $^3S_1$ & 1\,372\,998\,803\\
14 & $^3S_1$ & 1\,347\,874\,127 & 33 & $^3S_1$ & 1\,373\,265\,188\\
15 & $^3S_1$ & 1\,352\,673\,833 & 34 & $^3S_1$ & 1\,373\,505\,903\\
16 & $^3S_1$ & 1\,356\,377\,995 & 35 & $^3S_1$ & 1\,373\,724\,155\\
17 & $^3S_1$ & 1\,359\,296\,416 & 36 & $^3S_1$ & 1\,373\,922\,642\\
18 & $^3S_1$ & 1\,361\,636\,650 & 37 & $^3S_1$ & 1\,374\,103\,691\\
19 & $^3S_1$ & 1\,363\,541\,952 & 38 & $^3S_1$ & 1\,374\,269\,280\\
20 & $^3S_1$ & 1\,365\,113\,813 & 39 & $^3S_1$ & 1\,374\,421\,124\\
21 & $^3S_1$ & 1\,366\,425\,741 & 40 & $^3S_1$ & 1\,374\,560\,698\\
22 & $^3S_1$ & 1\,367\,532\,054 & 41 & $^3S_1$ & 1\,374\,689\,300\\
23 & $^3S_1$ & 1\,368\,473\,584 & 42 & $^3S_1$ & 1\,374\,808\,037\\
24 & $^3S_1$ & 1\,369\,281\,502 & 43 & $^3S_1$ & 1\,374\,917\,901\\
25 & $^3S_1$ & 1\,369\,979\,949 & 44 & $^3S_1$ & 1\,375\,019\,753\\
26 & $^3S_1$ & 1\,370\,587\,852 & 45 & $^3S_1$ & 1\,375\,114\,353\\
27 & $^3S_1$ & 1\,371\,120\,204 & 46 & $^3S_1$ & 1\,375\,202\,375\\
28 & $^3S_1$ & 1\,371\,589\,028 & 47 & $^3S_1$ & 1\,375\,284\,413\\
29 & $^3S_1$ & 1\,372\,004\,044 & 48 & $^3S_1$ & 1\,375\,360\,997\\
30 & $^3S_1$ & 1\,372\,373\,187 & 49 & $^3S_1$ & 1\,375\,432\,602\\
31 & $^3S_1$ & 1\,372\,702\,970 & 50 & $^3S_1$ & 1\,375\,499\,653\\
  \hline
  \end{tabular}
  \\
  \caption{Experimental value of the Rydberg state energies for the $5\mathrm{s}n\mathrm{s} \, ^3\mathrm{S}_1$ series. Listed here are the principal quantum number $n$, the Rydberg series and the experimental Rydberg state energy $E_{exp}$ expressed in MHz.  The uncertainty on these value is $10\,\mathrm{MHz}$, see in the text. }
  \label{tab:SseriesEnergyTable}
\end{table*}

\begin{table*}[!h]
 \small
 \center
  \begin{tabular}{|c c c  || c c c| }
  \hline 
  $n$ & Series & $E_{exp}$ (MHz) & $n$ & Series & $E_{exp}$ (MHz)\\ 
    \hline
12 & $^3D_1$ & 1\,344\,688\,300 & 31 & $^3D_1$ & 1\,372\,918\,946\\
12 & $^3D_2$ & 1\,344\,870\,880 & 31 & $^3D_2$ & 1\,372\,925\,192\\
13 & $^3D_1$ & 1\,349\,925\,421 & 32 & $^3D_1$ & 1\,373\,192\,657\\
13 & $^3D_2$ & 1\,350\,174\,107 & 32 & $^3D_2$ & 1\,373\,198\,097\\
14 & $^1D_2$ & 1\,353\,661\,576 & 33 & $^3D_1$ & 1\,373\,439\,862\\
14 & $^3D_1$ & 1\,353\,850\,897 & 33 & $^3D_2$ & 1\,373\,444\,629\\
14 & $^3D_2$ & 1\,354\,207\,372 & 34 & $^3D_1$ & 1\,373\,663\,874\\
15 & $^3D_1$ & 1\,356\,887\,051 & 34 & $^3D_2$ & 1\,373\,668\,084\\
15 & $^1D_2$ & 1\,356\,969\,526 & 35 & $^3D_1$ & 1\,373\,867\,493\\
15 & $^3D_2$ & 1\,357\,360\,134 & 35 & $^3D_2$ & 1\,373\,871\,228\\
16 & $^3D_1$ & 1\,359\,296\,415 & 36 & $^3D_1$ & 1\,374\,053\,114\\
16 & $^3D_2$ & 1\,359\,574\,504 & 36 & $^3D_2$ & 1\,374\,056\,446\\
16 & $^1D_2$ & 1\,359\,922\,783 & 37 & $^3D_1$ & 1\,374\,222\,798\\
17 & $^3D_1$ & 1\,361\,493\,566 & 37 & $^3D_2$ & 1\,374\,225\,784\\
17 & $^3D_2$ & 1\,361\,682\,770 & 38 & $^3D_1$ & 1\,374\,378\,312\\
18 & $^3D_1$ & 1\,363\,313\,169 & 38 & $^3D_2$ & 1\,374\,381\,002\\
18 & $^3D_2$ & 1\,363\,452\,486 & 39 & $^3D_1$ & 1\,374\,521\,191\\
19 & $^3D_1$ & 1\,364\,863\,538 & 39 & $^3D_2$ & 1\,374\,523\,622\\
19 & $^3D_2$ & 1\,364\,960\,612 & 40 & $^3D_1$ & 1\,374\,652\,762\\
20 & $^3D_1$ & 1\,366\,181\,658 & 40 & $^3D_2$ & 1\,374\,654\,968\\
20 & $^3D_2$ & 1\,366\,249\,880 & 41 & $^3D_1$ & 1\,374\,774\,191\\
21 & $^3D_1$ & 1\,367\,305\,009 & 41 & $^3D_2$ & 1\,374\,776\,201\\
21 & $^3D_2$ & 1\,367\,354\,291 & 42 & $^3D_1$ & 1\,374\,886\,489\\
22 & $^3D_1$ & 1\,368\,266\,930 & 42 & $^3D_2$ & 1\,374\,888\,324\\
22 & $^3D_2$ & 1\,368\,303\,621 & 43 & $^3D_1$ & 1\,374\,990\,551\\
23 & $^3D_1$ & 1\,369\,095\,323 & 43 & $^3D_2$ & 1\,374\,992\,230\\
23 & $^3D_2$ & 1\,369\,123\,407 & 44 & $^3D_1$ & 1\,375\,087\,158\\
24 & $^3D_1$ & 1\,369\,812\,958 & 44 & $^3D_2$ & 1\,375\,088\,706\\
24 & $^3D_2$ & 1\,369\,834\,979 & 45 & $^3D_1$ & 1\,375\,177\,012\\
25 & $^3D_1$ & 1\,370\,438\,257 & 45 & $^3D_2$ & 1\,375\,178\,436\\
25 & $^3D_2$ & 1\,370\,455\,884 & 46 & $^3D_1$ & 1\,375\,260\,722\\
26 & $^3D_1$ & 1\,370\,986\,135 & 46 & $^3D_2$ & 1\,375\,262\,039\\
26 & $^3D_2$ & 1\,371\,000\,499 & 47 & $^3D_1$ & 1\,375\,338\,834\\
27 & $^3D_1$ & 1\,371\,468\,700 & 47 & $^3D_2$ & 1\,375\,340\,055\\
27 & $^3D_2$ & 1\,371\,480\,583 & 48 & $^3D_1$ & 1\,375\,411\,845\\
28 & $^3D_1$ & 1\,371\,895\,834 & 48 & $^3D_2$ & 1\,375\,412\,978\\
28 & $^3D_2$ & 1\,371\,905\,792 & 49 & $^3D_1$ & 1\,375\,480\,182\\
29 & $^3D_1$ & 1\,372\,275\,642 & 49 & $^3D_2$ & 1\,375\,481\,234\\
29 & $^3D_2$ & 1\,372\,284\,085 & 50 & $^3D_1$ & 1\,375\,544\,238\\
30 & $^3D_1$ & 1\,372\,614\,826 & 50 & $^3D_2$ & 1\,375\,545\,219\\
30 & $^3D_2$ & 1\,372\,622\,055 & ~ & ~ & ~\\

  \hline
  \end{tabular}\\
  \caption{Experimental value of the Rydberg state energies for the $5\mathrm{s}n\mathrm{d} \, ^3\mathrm{D}_{1,2}$ and $5\mathrm{s}n\mathrm{d} \, ^1\mathrm{D}_{2}$ Rydberg series. Listed here are the principal quantum number $n$, the Rydberg series and the experimental Rydberg state energy $E_{exp}$ expressed in MHz.  The uncertainty on these value is $10\,\mathrm{MHz}$, see in the text. }
  \label{tab:DseriesEnergyTable}
\end{table*}


\bibliography{Rydberg_Spectro}

\end{document}